\documentclass[useAMS,usegraphicx,usenatbib]{mn2e}
\usepackage{graphicx}
\usepackage{subfigure}
\usepackage{url}
\usepackage{multirow}

\newcommand{\ks}{\rm\thinspace ks}

\newcommand{\keV}{\rm\thinspace keV}
\newcommand{\eV}{\rm\thinspace eV}

\newcommand{\rg}{\rm\thinspace $r_\mathrm{g}$}

\newcommand{\texttilde}{${}_{\textrm{\symbol{126}}}$}

\title[The emissivity profile of 1H\,0707-495]{Determination of the X-ray reflection emissivity profile of 1H\,0707-495}
\author[D. R. Wilkins \& A. C. Fabian]{D. R. Wilkins
  \thanks{E-mail: drw@ast.cam.ac.uk} 
and A. C. Fabian\\Institute of Astronomy, University of Cambridge, Madingley Road, Cambridge CB3 0HA}
\begin{document}

\date{Accepted 2011 February 2.  Received 2011 February 1; in original form 2010 October 18}

\pagerange{\pageref{firstpage}--\pageref{lastpage}} \pubyear{2010}

\maketitle

\label{firstpage}

\begin{abstract}
When considering the X-ray spectrum resulting from the reflection off the surface of accretion discs of AGN, it is necessary to account for the variation in reflected flux over the disc, \textit{i.e.} the emissivity profile. This will depend on factors including the location and geometry of the X-ray source and the disc characteristics. We directly obtain the emissivity profile of the disc from the observed spectrum by considering the reflection component as the sum of contributions from successive radii in the disc and fitting to find the relative weightings of these components in a relativistically-broadened emission line. This method has successfully recovered known emissivity profiles from synthetic spectra and is applied to XMM-Newton spectra of the Narrow Line Seyfert 1 galaxy 1H\,0707-495. The data imply a twice-broken power law form of the emissivity law with a steep profile in the inner regions of the disc (index 7.8) and then a flat region between 5.6\rg\ and 34.8\rg\ before tending to a constant index of 3.3 over the outer regions of the disc. The form of the observed emissivity profile is consistent with theoretical predictions, thus reinforcing the reflection interpretation.
\end{abstract}

\begin{keywords}
accretion discs -- black hole physics -- line: profiles -- X-rays: general.
\end{keywords}

\section[]{Introduction}

In the canonical `lamppost' model of AGN \citep{george_fabian}, hard X-rays originate from a source in the hot corona surrounding the central black hole and are observed as a power law continuum in the spectrum of the object (the so-called power law component, or PLC). X-rays from this source will also be reflected off the accretion disc, with photons being either Compton-scattered or undergoing photo-electric absorption which will lead to either Auger de-excitation or emission of a fluorescent line. This results in the so-called reflection-dominated component, or RDC of the spectrum. Within this component, a number of emission lines from species in the disc will be observed \citep{fabian+89}, as described by the self-consistent ionised reflector model \textsc{reflionx} \citep{ross_fabian}.

The reflected flux from the accretion disc (the \textit{emissivity profile}) varies over the disc resulting from the variation in flux received from the coronal X-ray source as a function of location on the disc. In the simplest case of a point source in flat, Euclidean spacetime, the emissivity at a point on the disc is proportional to the inverse-square of the distance from the source, multiplied by the cosine of the angle at which the ray hits the disc from the normal, giving a form $r^{-3}$ at large radius out from the source. In general relativity, however, rays will be focussed towards the black hole (and the inner disc) so naively one would expect a steeper fall-off in emissivity with distance from the black hole. The details of the emissivity profile will depend upon the location, spatial extent and geometry of the source.

When modelling the X-ray reflection spectrum from the accretion disc around the central black hole of an AGN, a (broken) power law emissivity profile ($\epsilon\propto r^{-\alpha}$), which describes the reflected power per unit area from the disc, is typically assumed and its parameters fit to the observed spectra. Power law forms of the emissivity profile are motivated by calculations tracing rays from coronal X-ray sources in the spacetime around the central black hole on to the accretion disc \citep{miniutti+03,dabrowski_lasenby,suebsuwong+06}. These calculations suggest a very steeply falling profile in the inner regions of the disc, then flattening off before tending to a constant power law slightly steeper than the $r^{-3}$ law one would expect in the classical case.

If the emissivity profile of an accretion disc can be determined from observations without \textit{a priori} assumption of its form, this will allow for testing of the light-bending hypothesis and could place constraints on the location and geometry of the coronal hard X-ray source.

We determine the emissivity profile of the accretion disc in the narrow line Seyfert 1 galaxy 1H\,0707-495, observed by \citet{fabian+09}. Previous attempts to recover the emissivity profile of the accretion disc of MCG--6-30-15 from observed spectra by \citet{dabrowski+97} fitted a free emissivity profile in the model of a relativistically broadened emission line, however here we recover the profile by considering directly the emission from each part of the accretion disc. The shape of a relativistically-broadened emission line, specifically the iron K$\alpha$ line, is considered as the sum of contributions from successive radii in the disc and by finding the relative weightings of these components, the emissivity profile is recovered.

\section[]{Relativistic Emission Line Profiles}

From a given region of the accretion disc, the frequency of emission of an emission line as measured by an observer at infinity will be altered from the rest-frame frequency by relativistic effects including Doppler shift, beaming and gravitational redshift, characterised by the transfer function \citep{laor-91}, defined such that the spectrum of the line seen by the observer at infinity is given by
\begin{equation}
	\label{linespectrum.equ}
	F_0(\nu_0) = \int I_e(r_e, \frac{\nu_0}{g})T(r_e,g)\,dg\,r_edr_e
\end{equation}
Where subscripts `$0$' denote quantities measured by the observer at infinity and `$e$' those measured in the rest frame of the emitter. The redshift parameter is $g = \frac{\nu_0}{\nu_e}$. For a line at frequency $\nu_e$, the rest-frame spectrum is simply 
\begin{equation}
I_e(r_e,\nu) = \delta(\nu - \nu_e)\epsilon(r_e)
\end{equation}
The emissivity profile, $\epsilon(r)$, gives the variation in reflected power as a function of radius, measured in the disc frame.

It is clear from the form (\ref{linespectrum.equ}) that the line profile can be considered as the sum of contributions from successive radii in the disc, with the number of photons received by the observer at infinity from an annulus of the accretion disc radius $r_e$ and thickness $dr_e$
\begin{equation}
	\label{photcount.equ}
	N_0(r_e, dr_e) = A'(r_e, dr_e)\epsilon(r_e) 
\end{equation}
Where $A'(r_e, dr_e)$ is the projected area of the annulus as seen by the observer, which from (\ref{linespectrum.equ}) is the classical area of the annulus $2\pi r_e dr_e$, multiplied by the integral of the transfer function over all frequencies (or over redshifts, $g$, after a change of variables), where the transfer function is defined to give photon counts per unit frequency, rather than fluxes (which is relevant here as when fitting to the reflection spectrum, the contribution from each region of the disc is given by its contribution to the total photon count).

\section[]{Emissivity Profile Determination}

\begin{figure}
	\centering
	\includegraphics[width=8cm]{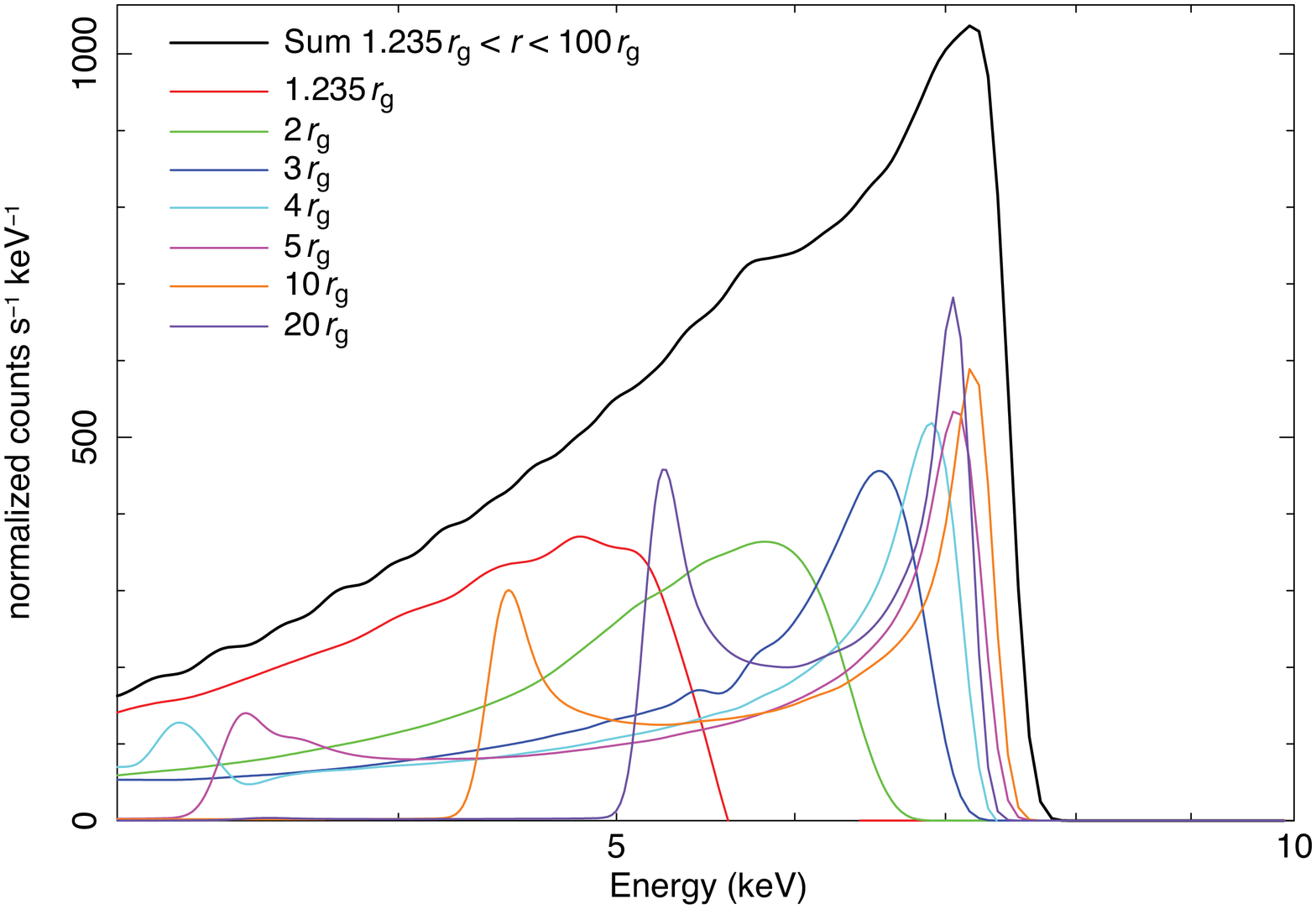}
	\caption{Relativistically broadened emission line profiles emitted from successive radii in the accretion disc ranging from 1.235\rg\ to 20\rg\ and the integrated line over the disc out to 100\rg, calculated using the \textsc{laor} line model.}
	\label{lineprofiles.fig}
\end{figure}

In order to determine the emissivity profile of the accretion disc from the observed X-ray spectrum of an AGN, the relativistically broadened iron K$\alpha$ emission line (which has a rest frame energy of around 6.4\keV, though varies slightly with the ionisation state of the iron) is considered, as above, to be the sum of contributions from successive radii in the disc. A relativistically broadened line will have a different profile if emitted from annuli of different radius, which can be seen in Fig.~\ref{lineprofiles.fig}, thus if the observed line can be broken down into contributions from different annuli and the relative weightings of these found, the reflected flux from each part of the disc (\textit{i.e.} the emissivity profile) can be found.

In the 3-10\keV\ region of the spectrum where the K$\alpha$ line lies, the spectrum is well described as the sum of a power law (the PLC) and the reflected line (RDC) which is described by the \textsc{reflionx} model, once convolved with the \textsc{kdblur} model to apply the relativstic blurring in the accretion disc. The spectrum is modelled in \textsc{xspec} as
\begin{equation}
	\mathrm{powerlaw} + \sum_{r_e} \mathrm{kdblur} \otimes \mathrm{reflionx}
\end{equation}
The radial intervals are selected to coincide with the radial binning of the transfer function used in the \textsc{laor} model (when fitting the model to the data, \textsc{xspec} looks up the transfer function from a pre-calculated table rather than evaluating it on-the-fly at each point, which would be computationally intensive) as it would be inappropriate to subdivide the disc further, since the resulting line profiles from these subdivided bins would be essentially the same as the same transfer function is used.

Only the 3-10\keV\ region of the spectrum is considered when determining the emissivity profile as this band contains a single emission line which can be decomposed into contributions identifying successive radii in the accretion disc according to the observed redshifts. When fitting for the emissivity, it is necessary to have a single emission line profile from which these components can be distinguished and as such the iron L region (0.5-1.0\keV), which contains multiple emission lines whose annular components will overlap once relativistic blurring is applied, is not suitable as emission from different regions of the disc cannot be distinguished here. Once the emissivity profile has been determined, however, it can be applied to a reflection spectrum covering the full energy range. Furthermore, the EPIC pn detector on board XMM-Newton has a spectral resolution of 150\eV\ (FWHM) at 6\keV\ meaning there are 46 energy bins in the 3-10\keV\ band but between 0.1 and 1\keV\, a resolution of 100\eV\ (FWHM) results in only nine bins, with only three before emission lines from oxygen appear below 0.65\keV, thus in the iron L band, there are insufficient spectral bins available to fit for the emissivity profile.

All model parameters are frozen to values which have been previously found to give the best fit for the source, except for the normalisations (photon counts) of the \textsc{reflionx} components, which are free parameters and are fit by minimising $\chi^2$ to find the relative weightings of the components. The inner and outer radius of each annulus is set in the blurring kernel \textsc{kdblur}. Each annulus has constant emissivity (the \texttt{index} parameter is set to zero) such that there is no variation in emitted flux across each annulus and the overall emissivity profile is given by the variation in photon counts between the annuli.

The projected area of each annulus as seen by the observer at infinity is found from the transfer function in the \textsc{laor} model (which is used by \textsc{kdblur} for the convolution), by summing the photon count over all frequencies at a given radius with a constant, flat emissivity. The emissivity profile is then calculated by dividing the normalisations of the \textsc{reflionx} components by these areas as per Equation \ref{photcount.equ}.

\subsection[]{Testing the Method}
\begin{figure*}
\centering
\begin{minipage}{170mm}
\subfigure[]{
\includegraphics[width=5.5cm]{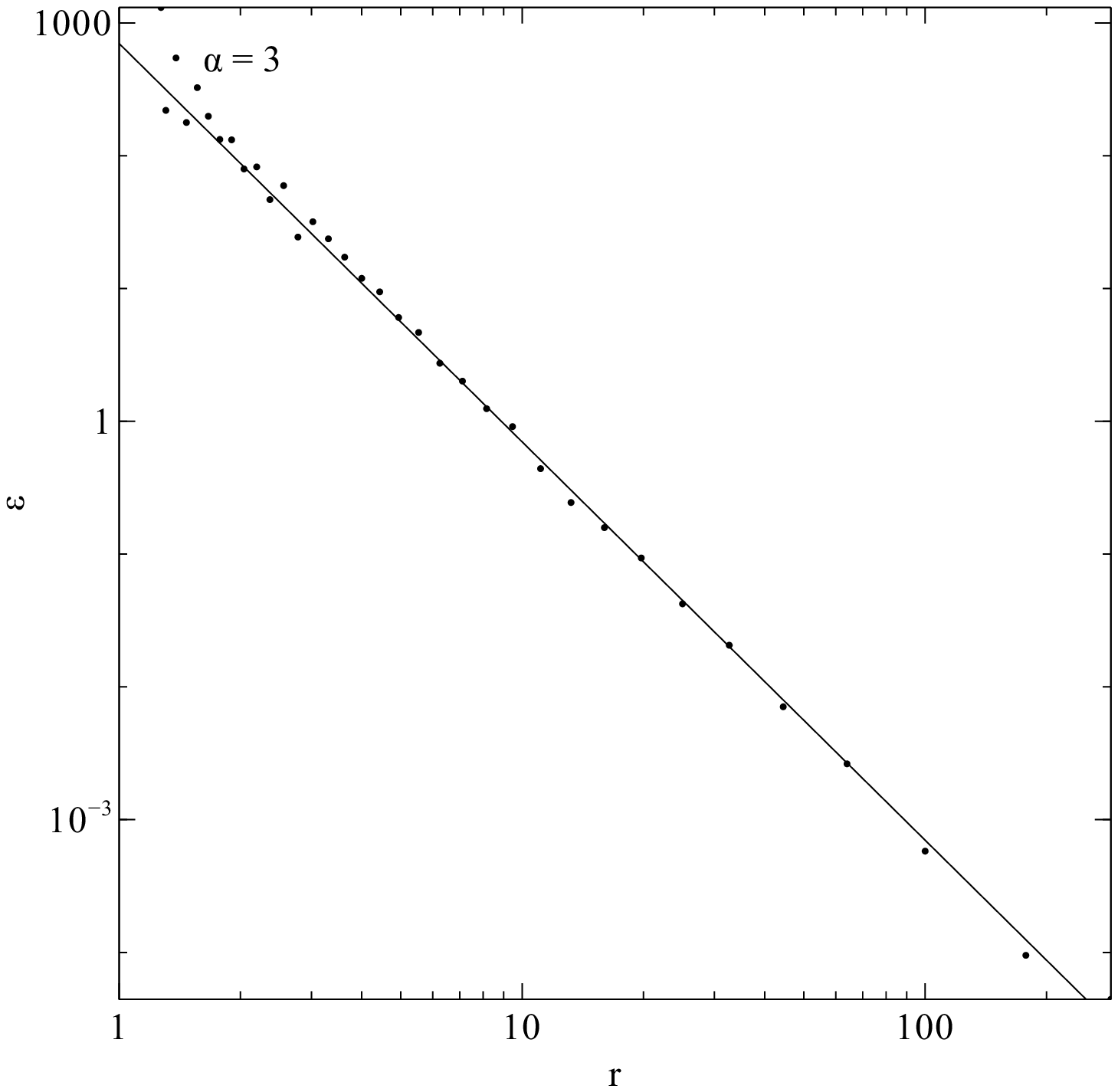}
\label{emis_test.fig:index3}
}
\subfigure[]{
\includegraphics[width=5.5cm]{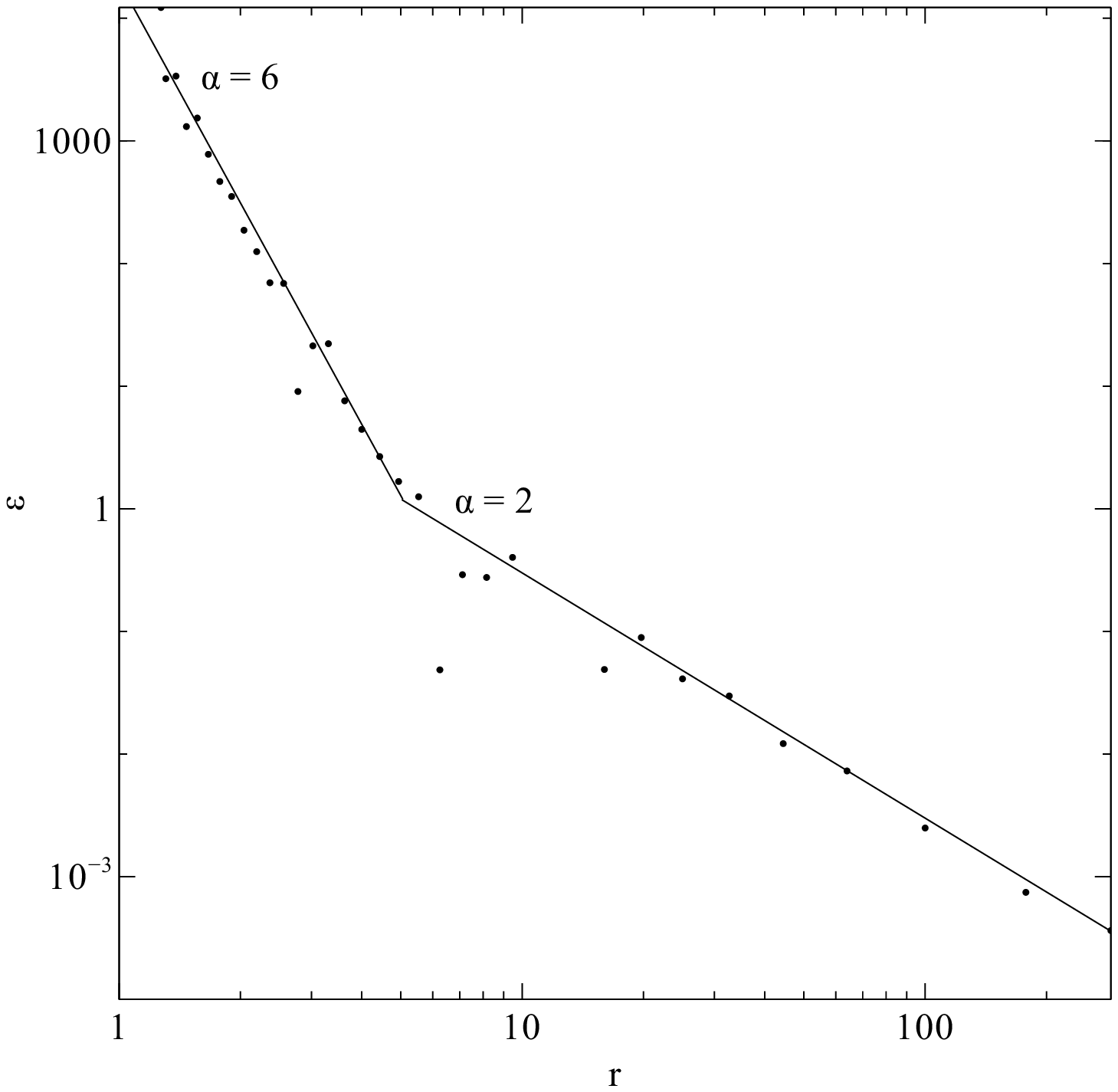}
\label{emis_test.fig:broken}
}
\subfigure[]{
\includegraphics[width=5.5cm]{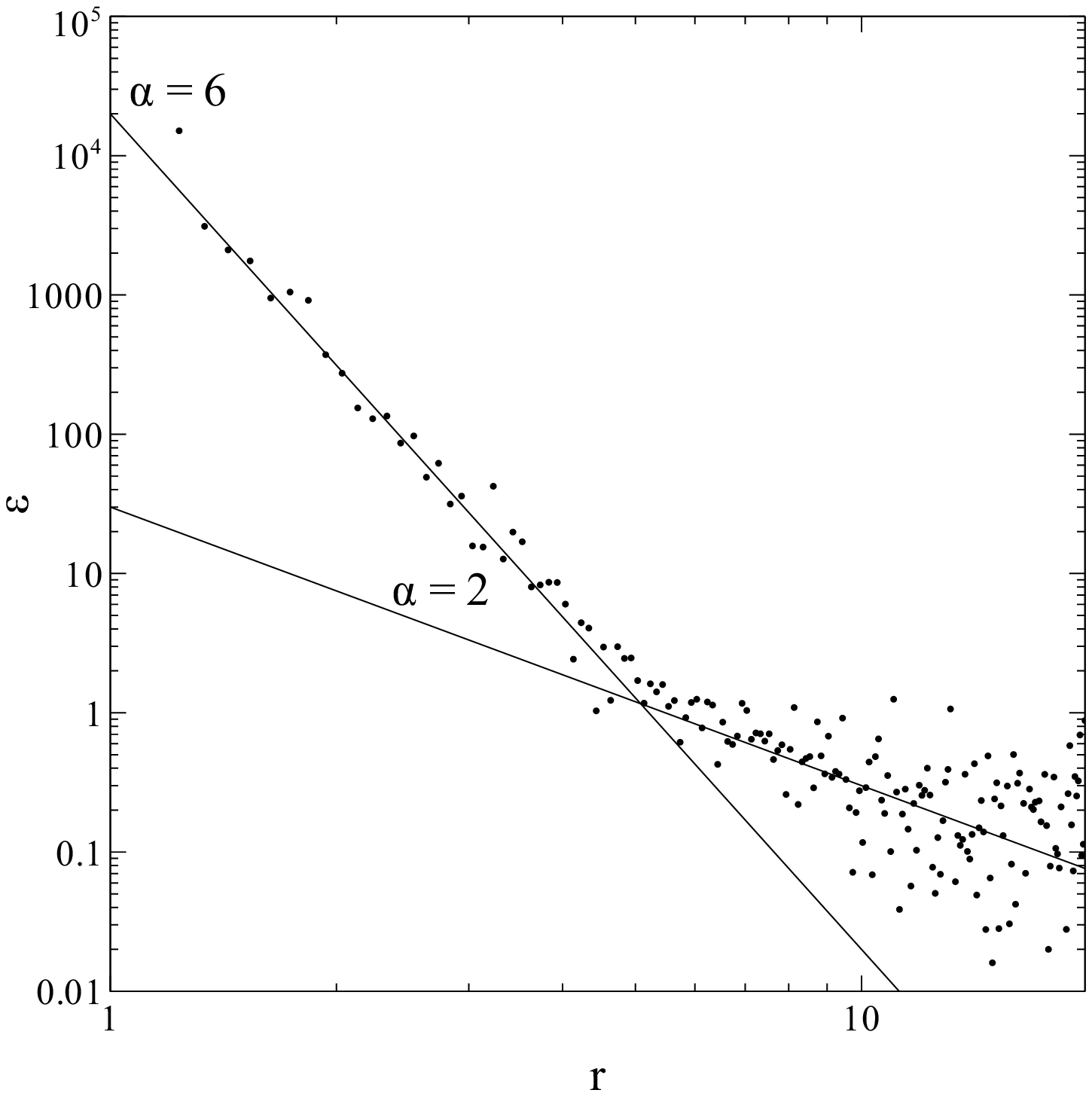}
\label{emis_test.fig:broken_scatter}
}
\caption[]{Recovered emissivity profiles from synthetic spectra to test the method, each consisting of a single emission line with \subref{emis_test.fig:index3} a single power law emissivity profile, index 3 and \subref{emis_test.fig:broken} a once-broken power law emissivity profile, indices 6 and 2 with the break at a radius of 5\rg. \subref{emis_test.fig:broken_scatter} is for the same profile as \subref{emis_test.fig:broken} but illustrates the scatter that occurs in regions where the radial points are more finely spaced than the radial bins in the \textsc{laor} model.}
\label{emis_test.fig}
\end{minipage}
\end{figure*}

It is not immediately obvious that a given emissivity profile will lead to a unique form of the observed spectrum, particularly once folded through the response function of a real X-ray detector, since the contributions to the line from different radii are not, strictly speaking, orthogonal functions. Before applying the method to real data, it is therefore important to carry out preliminary tests, applying it to synthetic spectra whose emissivity profiles are known and checking that these profiles can be recovered correctly.

This method has been applied to synthetic spectra in which line profiles with known emissivity profiles, formed from the \textsc{laor} emission line model, were folded through the response matrix of the EPIC pn detector on board XMM-Newton, using the \texttt{fakeit} command in \textsc{xspec} to simulate the spectra that would be observed if sources emitting such lines were observed with this instrument. Synthetic spectra were calculated for an exposure time of 300\ks\ and the incident flux in the emission line was normalised to unity, since here we are interested only in the overall shape of the line spectra with no background. The response matrix of the detector maps only incident energies onto detector channels. It is independent of the incident flux which will just provide an overall scaling that is irrelevant here since we are interested in whether the contributions from successive annuli can be recovered from the overall shape of the spectrum.

A variety of line profiles were tested, with single power law emissivity profiles with indices ranging from 2.0 to 6.0. In addition to this, once-broken power law profiles were tested (these model line spectra are created using the \textsc{laor2} variant of the model). It can be seen from Fig.~\ref{emis_test.fig} that these profiles were successfully recovered.

Scatter at larger radius appears where the radial step is finer than the binning used in the \textsc{laor} line model where the transfer function is evaluated. As a result, the line profiles from neighbouring radii in the same bin are essentially the same. They cannot be distinguished in the fit, leading to multiple points with essentially equal weighting rather than following the true emissivity profile as illustrated in Figure \ref{emis_test.fig:broken_scatter}. This scatter is reduced when the radial interval in the fit is chosen to coincide with the radial binning of the \textsc{laor} transfer function.

\section[]{Emissivity Profile of 1H\,0707-495}

The emissivity profile of the accretion disc in this Narrow Line Seyfert 1 galaxy was determined using the above method applied to the spectrum obtained using the EPIC pn detector on board XMM-Newton during four consecutive orbits with an exposure time of 500\ks\ \citep{zoghbi+09}. The values of the frozen parameters (\textit{i.e.} those other than the photon counts from the reflection components) were taken from the best fit model of \citet{zoghbi+09} and are shown in Table \ref{par.tab}.

\begin{table}

\caption{Values of frozen model parameters used in emssivity profile determination.}
\begin{tabular}{lll}
  	\hline
   	\textbf{Component} & \textbf{Parameter} & \textbf{Value} \\
	\hline
	powerlaw (PLC) & Photon index, $\Gamma$ & 3.09 \\
	& Photon count, \textit{norm} & $1.50\times 10^{-3}$ \\
	\hline
	kdblur (RDC) & Inclination, $i$ & $53.96^\circ$ \\
	\hline
	reflionx (RDC) & Photon index, $\Gamma$ & = \texttt{powerlaw:}$\Gamma$ \\
	& Iron abundance / solar & 8.88 \\
	& Ionisation parameter, $\xi$ & 53.44 \\
	& Redshift, $z$ & $4.10\times 10^{-2}$ \\
	\hline
\end{tabular}
\label{par.tab}
\end{table}

\begin{figure}
	\centering
	\includegraphics[width=5.5cm,angle=270]{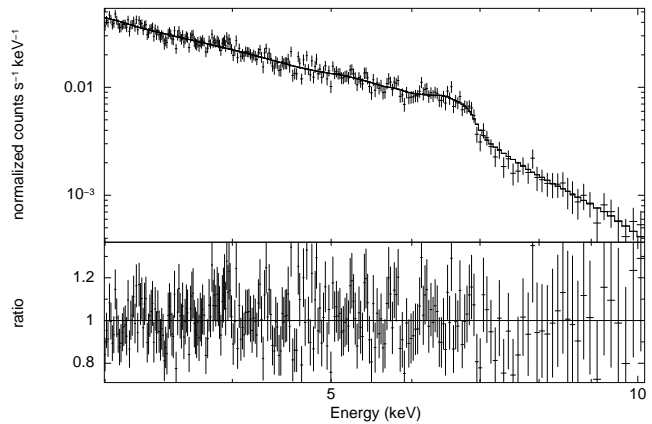}
	\caption{Fit to the iron K$\alpha$ edge (3-10\keV) of 1H\,0707-495 with the disc reflection considered as the sum of components from successive radii. $\chi^2/\mathrm{DoF} = 255.48 / 227 = 1.1255$.}
	\label{klinefit.fig}
\end{figure}

The normalisations of the reflection components were initially fit in the energy range 3-10\keV, which encompasses the entire iron K$\alpha$ line. The resulting fit is shown in Fig.~\ref{klinefit.fig} and gives $\chi^2/\mathrm{DoF} = 255.48 / 227 = 1.1255$, indicating that the model describes the data well and that there are no significant components other than the power law continuum and reflection from the disc contributing to the spectrum in this energy range.

\begin{figure*}
\centering
\begin{minipage}{170mm}
\subfigure[3-10\keV]{
\includegraphics[width=8cm]{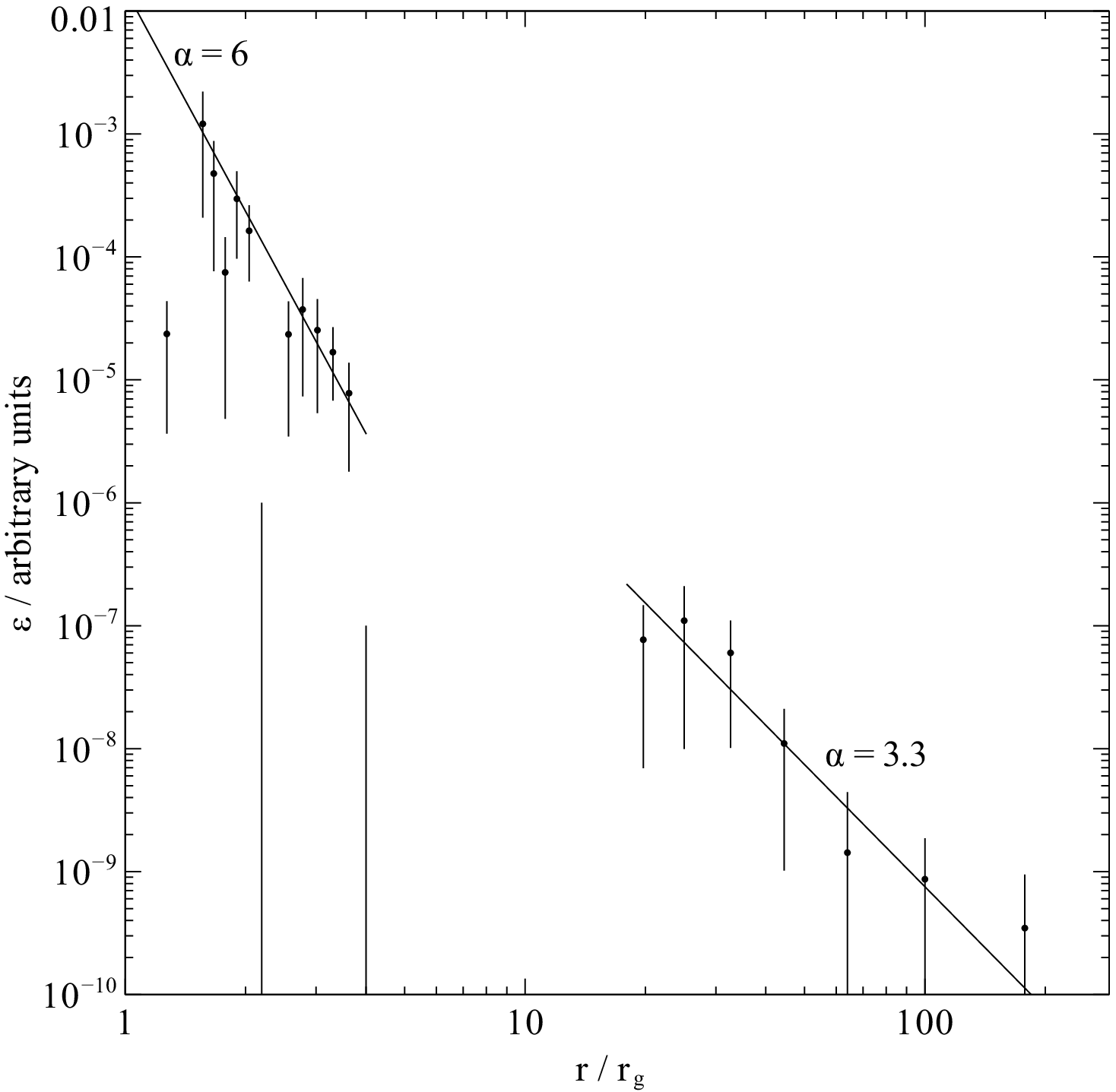}
\label{1h0707_emis.fig:3-10kev}
}
\subfigure[3-5\keV]{
\includegraphics[width=8cm]{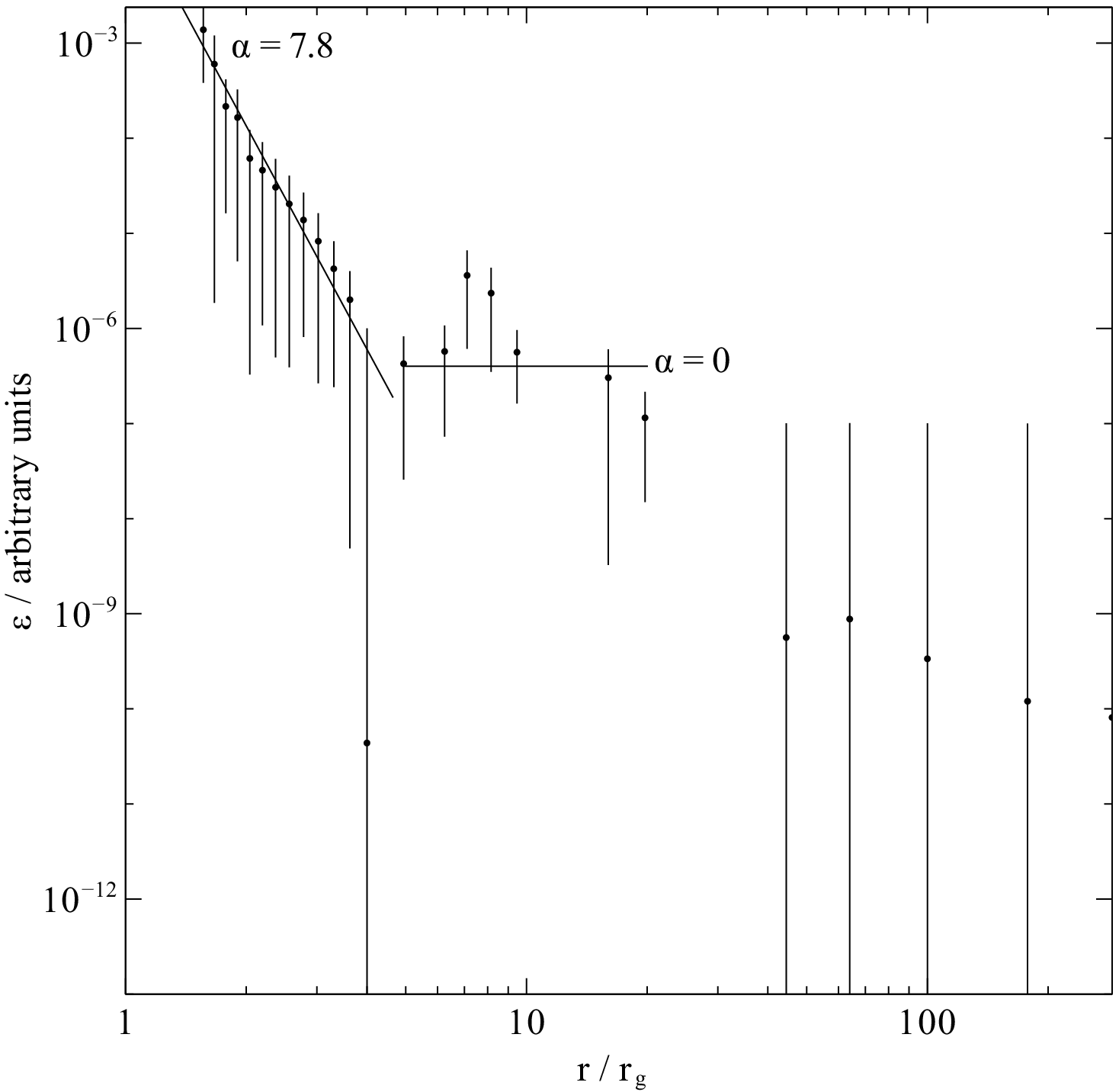}
\label{1h0707_emis.fig:3-5kev}
}
\caption[]{Emissivity profile of 1H\,0707-495 recovered from fitting to the spectrum in the energy ranges \subref{1h0707_emis.fig:3-10kev} 3-10\keV\ and \subref{1h0707_emis.fig:3-5kev} 3-5\keV. Error bars show $1\sigma$ confidence limits corresponding to $\Delta\chi^2=1$. Lines show power laws of index 6.0 and 3.3 in \subref{1h0707_emis.fig:3-10kev} and 7.8 for the inner disc as well as a flat profile, for reference, through the middle region in \subref{1h0707_emis.fig:3-5kev}. In the 3-5\keV\ fit, error bars are large at radii beyond 30\rg\ as emission in the iron K$\alpha$ line from these radii is excluded from the fit range, so the points are unconstrained.}
\label{1h0707_emis.fig}
\end{minipage}
\end{figure*}

The emissivity profile found from this fit by dividing the best-fit \textsc{reflionx} normalisations by the projected areas of the annuli is shown in Fig.~\ref{1h0707_emis.fig:3-10kev}.

The model was also fit to the spectrum in the restricted energy range 3-5\keV. This energy range includes contributions to the 6.4\keV\ emission line from regions of the disc out to 20\rg, however contributions to the line emitted from radii beyond this are mostly excluded, with only their redshifted wings included. Since the majority of the line flux is in the energy range above this and originates from the outer disc, the fit over the full line will be weighted heavily towards these radii, so fitting to the restricted 3-5\keV\ range will give better detail of the inner regions of the disc. The emissivity profile determined from the results of this restricted fit is shown in Fig.~\ref{1h0707_emis.fig:3-5kev}.

From the above fits, a steep emissivity profile is apparent in the inner regions of the disc, with the 3-10\keV\ fit indicating a power law index of 6.0 out to a radius of around 3\rg\ and tending to a value of around 3.3 at large radius.

In the 3-10\keV\ fit, the points in the region 4-20\rg\ are missing, with the best fit values to these normalisations being zero. However the $1\sigma$ error limits on these points are very large (on the arbitrary scale in Fig.~\ref{1h0707_emis.fig:3-10kev}, these points could be as high as $10^{-6}$), showing that the photon counts from this region of the disc are poorly constrained by these data. This is most likely due to the $\chi^2$ statistic in a fit over this energy range being dominated by the line flux from the outer regions of the disc which is much greater than that received from the inner region. The line profiles from the innermost regions out to 4\rg\ account for the shape of the line at low energy (and have a relatively high photon count due to the apparent steepness of the emissivity profile). Once these contributions and those from the outer disc are accounted for, the 4-20\rg\ region corresponding to the middle band of the emission line is not well constrained. This is remedied in the 3-5\keV\ fit where the outer region of the disc (outside 20\rg) which dominated the line flux is removed and more detail of the inner disc is seen.

The fit in the region 3-5\keV\ steepens the index of the emissivity profile to around 7.8 out to a radius of 5\rg. This fit also indicates a flattening of the emissivity profile in the region 5-30\rg\ to an index of between 0 and 1, however the error bars in this region are large and these points do not appear in the 3-10keV fit because the contributions from these regions lie in the middle of the line profile so are masked by the contributions from either side. Note that in this fit, the points in the outer disc are poorly constrained and have large error bars because the energy range of their contributions to the line is omitted from the fit.

\begin{figure}
	\centering
	\includegraphics[width=85mm]{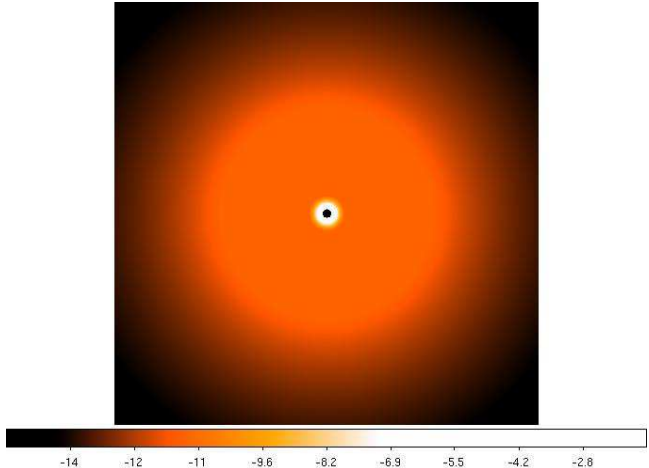}
	\caption{Pseudo-image of the inner regions of the accretion flow in 1H\,0707-495 out to 60\rg\ formed from the observed emissivity profile. Shading corresponds to reflected flux in the disc frame on a logarithmic scale.}
	\label{1h0707_disc.fig}
\end{figure}

The determined emissivity profile can be used to construct a pseudo-image of the accretion flow in 1H\,0707-495, showing the reflected flux (measured in the rest frame of the disc) as a function of position in the disc (Figure \ref{1h0707_disc.fig}).

While this analysis was completed using the \textsc{kdblur} model which convolves the reflection spectrum with the \textsc{laor} line model for efficiency of computation, the analysis can be completed using other relativistic disc convolution models. One such model is the \textsc{relconv} model of \citet{dauser+10} which convolves the reflection spectrum with the \textsc{relline} profile, providing finer radial binning than the \textsc{laor} model. Using \textsc{relconv} yields very much the same result as is obtained here. A couple of further points are obtained in the flat 5-30\rg\ region of the emissivity profile when performing the fit over the full 3-10\keV\ range consistent with the above result, however this region is still largely unconstrained (until the 5-10\keV\ band is excluded from the fit). This indicates that this is not due to systematics in the \textsc{laor} model but is rather the masking of this region by the high flux from the outer disc in the observed spectrum.

\section[]{Testing the Result}

It is important to test these results for the emissivity profile, in particular the flattened region of the profile at 5-30\rg, by fitting one continuous reflection spectrum with the appropriate emissivity profile over the entire disc. It would appear from the above that the closest analytic form for the profile is a twice-broken power law.

In \textsc{xspec}, the emissivity profile of the disc is applied by the relativistic blurring kernel, \textsc{kdblur} (single power law emissivity) or \textsc{kdblur2} (once-broken power law), to the rest-frame reflection spectrum \textsc{reflionx}. The blurred spectrum is obtained by convolving the original spectrum with the profile of a relativistically broadened line (delta-function spectrum in the rest frame) from the \textsc{laor} or \textsc{laor2} models.

Modified versions of these models were written to give a relativistically blurred reflection spectrum from the disc corresponding to a twice-broken power law emissivity profile: The model \textsc{kdblur3}, which provides the blurring kernel based upon the line profile \textsc{laor3}\footnote{http://www-xray.ast.cam.ac.uk/\texttilde drw/models/kdblur3}, with which \textsc{xspec} may convolve the rest frame spectrum. These models have the three power law indices and two break radii for the emissivity profile as input parameters.

The spectrum was fit in the region 3-10\keV\ (the iron K$\alpha$ line, with no significant spectral components other than the power law continuum and reflection component) with the model
\begin{equation}
	\mathrm{powerlaw} + \mathrm{kdblur3} \otimes \mathrm{reflionx}
\end{equation}
with the power law indices and break radii of the emissivity profile fit as free parameters, within limits to constrain them to roughly the values observed (\textit{i.e.} a steeper index tending to a moderate index, with a flat profile in the middle region) but sufficiently broad not to place any particular bias on the results. Other parameters of the models were again set to the best-fit values of \citet{zoghbi+09}. The results of this fit are shown in Table \ref{kdblur3fit.tab}.

Table \ref{kdblur2kdblur3.tab} shows this fit compared to an equivalent fit using only a once-broken power law as had been assumed in previous work in the analysis of the spectra.

\begin{table}

\caption{Fit to the emissivity profile of 1H\,0707-495 (3-10\keV) using the \textsc{kdblur3} model (twice-broken power law emissivity).}
\begin{tabular}{llll}
  	\hline
   	\textbf{Parameter} & \textbf{Fit Range} & \textbf{Value} & \textbf{Error} ($1\sigma$) \\
	\hline
	Index 1 	& 	2 -- 12	 & 	7.83	 &	$-0.66,+3.97$\\
	Break radius 1 	& 	1.235 -- 10\rg 	&	5.60\rg	& $-0.48,+0.15$	\\
	Index 2 	&	0 -- 2	&	$7.84\times 10^{-5}$	&	$+0.38$   \\
	Break radius 2	&	5 -- 400\rg	&	34.75\rg	& $-4.72,+4.42$\\
	Index 3		&	1.5 -- 10	&	3.30	&	$-0.32,+0.43$ \\
	\hline
\end{tabular}
\label{kdblur3fit.tab}
\end{table}

\begin{table}
\centering

\begin{tabular}{lll}
  	\hline
   	& \textbf{kdblur2} & \textbf{kdblur3} \\
	\hline
	Index 1 	& 	4.82	 & 	7.83 \\
	Break radius 1 	& 	6.85\rg 	&	5.60\rg	\\
	Index 2 	&	2.09	&	$7.84\times 10^{-5}$	\\
	Break radius 2	&	---	&	34.75\rg \\
	Index 3		&	---	&	3.30	\\
	\hline
	$\chi^2$ & 287.52 & 264.32 \\
	$\chi^2/\mathrm{DoF}$ & 1.1144 & 1.0365 \\
	\hline
\end{tabular}
\caption{Comparison between fits to the iron K$\alpha$ line (3-10\keV) of 1H\,0707-495 using models with once- and twice-broken power law emissivity profiles (\textsc{kdblur2} and \textsc{kdblur3}, respectively).}
\label{kdblur2kdblur3.tab}
\end{table}

\begin{figure}
	\centering
	\includegraphics[height=85mm,angle=270]{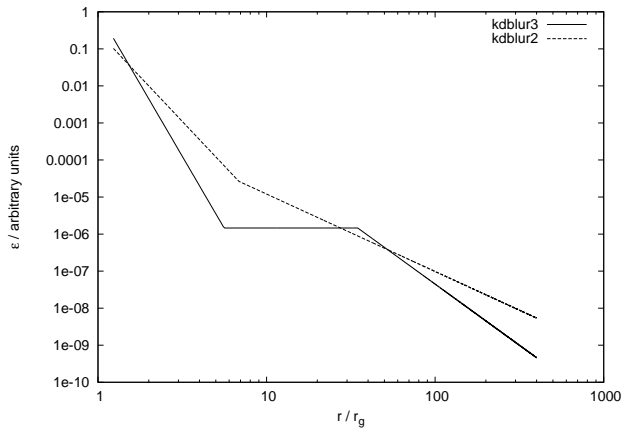}
	\caption{Comparison between the best fit emissivity profiles obtained for 1H\,0707-495 using the \textsc{kdblur3} ($\chi^2/\mathrm{DoF} = 1.037$) and \textsc{kdblur2} ($\chi^2/\mathrm{DoF} = 1.114$) models.}
	\label{kdblur3kdblur2.fig}
\end{figure}

Modelling the emissivity profile as a twice-broken power law fits the data substantially better than using a once-broken power law, giving a reduction in $\chi^2$ of 23.2 (a reduction from 1.11 per degree of freedom to 1.04 per degree of freedom). The best-fit parameters obtained using the \textsc{kdblur3} model agree well with the emissivity profiles recovered from the sum of reflection components, with a power law index of 7.8 in the inner disc out to 5\rg\ and flattening to a very low index of $8\times 10^{-5}$ from here to 35\rg\ before tending to an index of 3.3 over the outer regions of the disc. The best fit emissivity profiles obtained using the \textsc{kdblur3} and \textsc{kdblur2} models are shown in Fig. \ref{kdblur3kdblur2.fig}.

The parameters are well constrained to these values by the $1\sigma$ confidence limits, with the exception of the index in the middle, flattened region, where the lower limit is not well constrained by the data but an upper limit of 0.4 is placed on the index here. Likewise, the lower limit of the power law index for the inner disc is found, yet the upper bound is less well constrained with a confidence limit of $+3.97$. This is likely due to the low flux received from the inner disc since photons originating from here are more likely to fall into the black hole, thus increasing the emission from the inner disc by steepening the emissivity profile will have less of an effect on the line profile detected at infinity.

\begin{figure*}
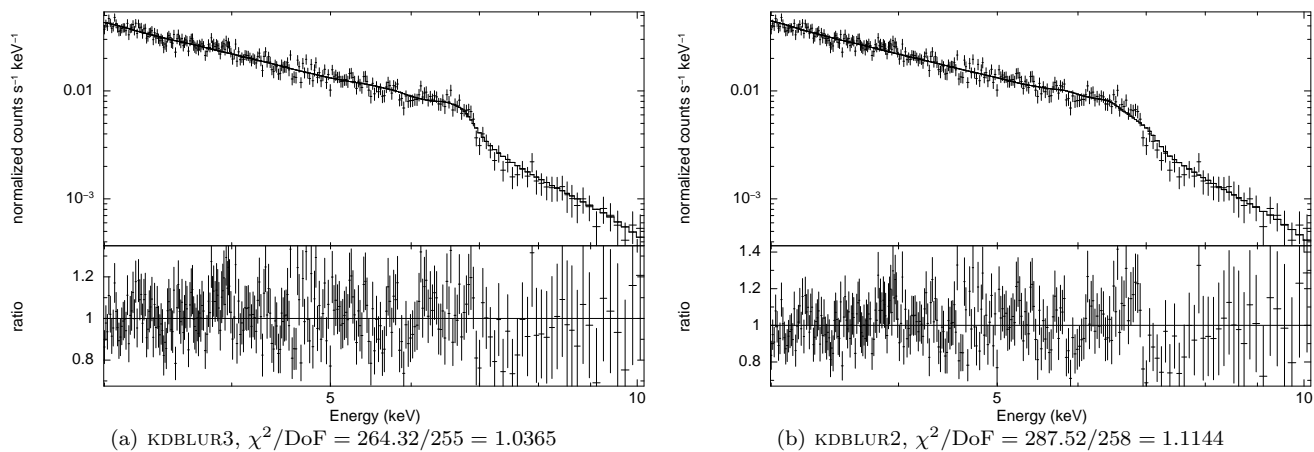

\centering
\begin{minipage}{170mm}
\subfigure[\textsc{kdblur3}, $\chi^2/\mathrm{DoF} = 264.32 / 255 = 1.0365$]{
\includegraphics[width=55mm,angle=270]{1h0707_kdblur3_3-10kev.ps}
\label{kdblurfits.fig:kdblur3}
}
\subfigure[\textsc{kdblur2}, $\chi^2/\mathrm{DoF} = 287.52 / 258 = 1.1144$]{
\includegraphics[width=55mm,angle=270]{1h0707_kdblur2_3-10kev.ps}
\label{kdblurfits.fig:kdblur2}
}
\label{fig:subfigureExample}
\caption[]{Best fits to the iron K$\alpha$ edge of 1H\,0707-495 using \subref{kdblurfits.fig:kdblur3} twice-broken and \subref{kdblurfits.fig:kdblur2} once-broken power law emissivity profiles (\textsc{kdblur3} and \textsc{kdblur2} respectively).}
\label{kdblurfits.fig}
\end{minipage}
\end{figure*}

The fits to the spectrum using the two models are shown in Fig.~\ref{kdblurfits.fig}. The form of the edge between 6.5 and 7\keV\ is reproduced much better by the \textsc{kdblur3} model than by \textsc{kdblur2}.

\subsection{Further Spectral Components}

Adding a narrow (Gaussian) line at 6.4\keV\ (fitting for the line width, $\sigma$, and photon count) to account for X-ray emission from a more distant reflector from which relativistic blurring is not seen does not significantly improve the fit to the data. The reduced $\chi^2$ changes to only 1.01 (compared with 1.03) when the emissivity profile is frozen at the values determined above. The flux in the narrow line is 100 times less than that seen in the broad line from the disc, indicating that this component is not significant. Allowing the parameters of the \textsc{kblur3} emissivity profile to vary also in the fit yields a similar result with the observed emissivity profile not changing greatly (it is steepend slightly over the outer disc to compensate for the extra flux in the narrow line coinciding with energies from the outer disc, but the fit is not improved with a reduced $\chi^2$ of 1.03).

Adding a second ionisation component in the disc (either simply adding a second component that is cospatial with the first, or dividing the disc into two ionisation components for the inner and outer disc and fitting for the break point) does not improve the fit. The reduced $\chi^2$ statistic is unchanged at 1.03 when the disc from 45\rg\ outwards has the ionisation parameter reduced to $\xi = 28$ (while that in the inner disc is still found to be around $\xi = 53$). The best fit emissivity profile for the disc with two ionisation components is found to be essentially unchanged from the previous result using one component.

\section[]{Discussion}

It is evident from the results here that the emissivity profile of 1H\,0707-495 resembles the form predicted by \citet{miniutti+03}, with a steep power law ($\epsilon\propto r^{-\alpha}$) in the inner regions of the disc, found to have an index of around 7.8, flattening to a very shallow index, close to zero, before tending to a constant index of 3.3, slightly steeper than the Euclidean case, at large radius.

\begin{figure}
	\centering
	\includegraphics[height=85mm,angle=270]{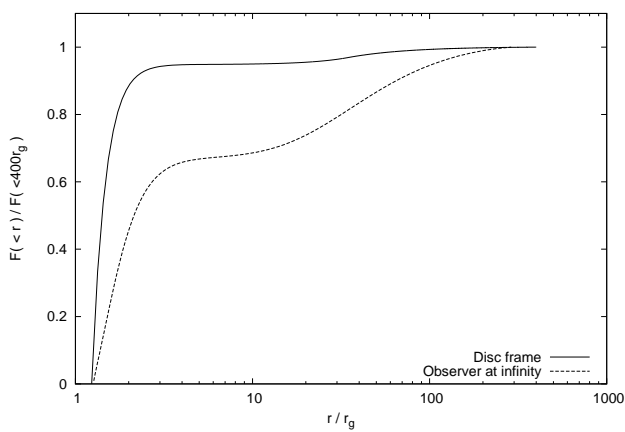}
	\caption{Cumulative reflected X-ray flux distribution, $F(<r)$, as a function of radius in the disc as implied by the determined emissivity profile for 1H\,0707-495 measured in the disc frame compared with that as measured by an observer at infinity (taking into account the transport of radiation from the disc to the observer around the black hole for 1H\,0707-495 which is observed at an inclination angle of $53^\circ$).}
	\label{emis_integral.fig}
\end{figure}

Such an emissivity profile suggests that 90\% of the X-ray flux reflected from the accretion disc (where $F(<r) = \int_{r_\mathrm{in}}^r \epsilon(r')\,r'dr'$, measured in the local disc frame) is reflected from the innermost regions of the disc, within 2\rg\ of the central black hole as shown in Fig. \ref{emis_integral.fig} (the solid line shows the cumulative flux distribution measured in the disc frame and the dashed line that measured by an observer at infinity where the outer disc is more dominant due to the larger solid angle subtended and the tendency of photons emitted from the inner regions of the disc to fall into the black hole --- as seen at infinity, only around 66\% of the reflected flux originates from the inner disc, within 4\rg).

Qualitatively, this form of the emissivity profile can be understood by considering an isotropic point source above the disc in a flat, Euclidean spacetime. The flux received from the source at each point on the disc, and thus the reflected power (emissivity) from that point will vary simply as the inverse square of the distance from the primary X-ray source, projected into the direction normal to the disc plane, \textit{i.e.} if the source is at a height $h$ above $r=0$ in the disc, the emissivity will go as $(r^2 + h^2)^{-1}\cos\vartheta$, with $\cos\vartheta=\frac{h}{\sqrt{r^2+h^2}}$, the angle from the normal at which the ray hits the disc. This will be constant (a flat profile) in the limit $r \ll h$ (the inner disc) and tending to $r^{-3}$ at large radius, giving a power law emissivity profile with index $3$.

In the presence of the black hole, gravitational light-bending will act to focus light rays from the point source towards the black hole and will increase the flux incident on the inner regions of the disc relative to the outer regions. This will cause the emissivity to fall off more steeply in the central regions of the accretion disc as is observed with 1H\,0707-495, before reaching a power law index of around $3.3$ at large radius, slightly steepened from the classical value of of $3.0$.

In addition to gravitational light bending focussing the light rays on to the disc, the emissivity profile will be affected by relativistic effects on the disc itself. While the classical area of an annulus radius $r$ and thickness $dr$ is $2\pi r dr$, the proper area as measured in the rest frame of the disc for a given $dr$ is increased as $r$ decreases, as the space is warped close to the black hole and proper radial distances increase at small $r$ as described by the Kerr metric. As the disc orbits the black hole (which on the last stable orbit will be at half the speed of light), the disc material will be length-contracted by a factor of $\gamma$, the Lorentz factor according to a stationary observer, so the area of the radial bins is increased further in the disc frame by a factor of $\gamma$. These effects increase the proper area of a radial bin closer to the black hole so will tend to reduce the steepness of the emissivity profile (as the incident flux is divided by the disc area).

The dominant effect steepening the emissivity profile close to the black hole is dilation of the proper time in the disc frame (\textit{i.e.} gravitational redshift). If the X-ray source emits photons at a constant rate in its own rest frame, the arrival rate in the disc frame will be enhanced if that disc element is closer to the black hole, as its proper time elapses more slowly. To an approximation, the photon arrival rate varies between different regions of the disc like $\dot{t}(r)$, the derivative of co-ordinate time with respect to the disc element's proper time, which increases steeply moving closer to the black hole. Furthermore, since emissivity is defined as the \emph{flux} emitted, a further factor of approximately $\dot{t}$ is introduced as each photon is blueshifted as it travels toward the black hole.

If the X-ray source is moving (\textit{e.g.} orbiting the rotation axis), emission will be relativistically `beamed' into the direction of motion of the source. This will enhance emission in front of the source while decreasing emission behind and to the sides of the direction of motion. If the source is orbiting the rotation axis on a locus above the disc plane, emission will be enhanced on to regions of the disc below the orbital locus, while the emissivity outside the orbit is reduced. If the source is close to the black hole, this will further steepen the emissivity profile.

\begin{figure}
	\centering
	\includegraphics[width=85mm]{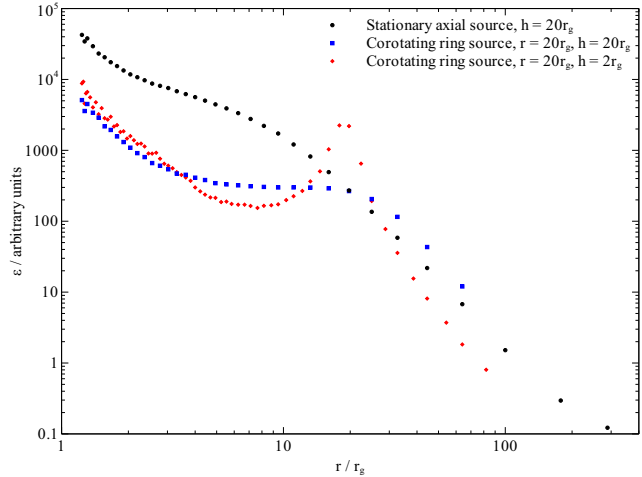}
	\caption{Theoretical accretion disc emissivity profiles due to (a, black) a stationary point source on the rotation axis above the disc plane, (b, blue) and (c, red)  ring sources co-rotating with the disc below, from ray tracing simulations, taking into account relativistic effects on the area of the accretion disc. The emissivity profile steepens over the innermost regions of the disc to a power law index of up to 7 in (a) and 5 in (b) then flattens off before tending to a constant power law index of around 3 over the outer disc. Increasing the radius of the ring source moves the break-point between the flat region and the outer disc outwards to the point approximately below the source. (c) illustrates the extreme case of a source at large radius while at a low height above the disc, enhancing the steepening over the inner regions of the disc. The emissivity here rises directly below the source with a steep fall-off at large radius since due to the low height, many photons are intercepted by the disc before they can propagate further.}
	\label{emis_theoretical.fig}
\end{figure}

The combination of these effects is illustrated by theoretical emissivity profiles created by a code written to trace rays from an isotropic point source in the corona above the plane of the accretion disc and counting the photons incident on radial bins in the equatorial plane, dividing by the area of the bins while taking into account the above effects (Fig.~\ref{emis_theoretical.fig}). Such simple simulations can qualitatively explain the form of the observed emissivity profile.

The best fit to the spectrum with reflection components originating from successive radii indicates a non-zero contribution to the spectrum from photons originating from the inner regions of the disc right down to the innermost stable orbit for a maximally spinning Kerr black hole at 1.235\rg. This suggests that the accretion disc extends down to the innermost stable orbit. The contribution received from the innermost regions of the accretion disc by an observer at infinity is, however, small once the projected area of the annuli (including the effect of the propagation of rays around the black hole to the observer) is accounted for. As such, the drop in the observed emissivity over several decades across the innermost regions may not be as significant to the observed line profile as first it seems. 

The steep power law emissivity profile determined here without \textit{a priori} assumption of its form and the agreement of this form with theoretical emissivity profiles calculated for point sources around the black hole show that the steep emissivity profiles previously required to explain the emission seen in terms of reflection of X-rays from a hard X-ray source near the black hole off of the accretion disc are indeed plausible. The reflection spectrum has a significant contribution from the inner regions of the disc right the way down to the inner-most stable orbit.

\begin{figure}
\centering
\subfigure[$r_\mathrm{ISCO} < r < 400r_\mathrm{g}$]{
\includegraphics[width=8cm]{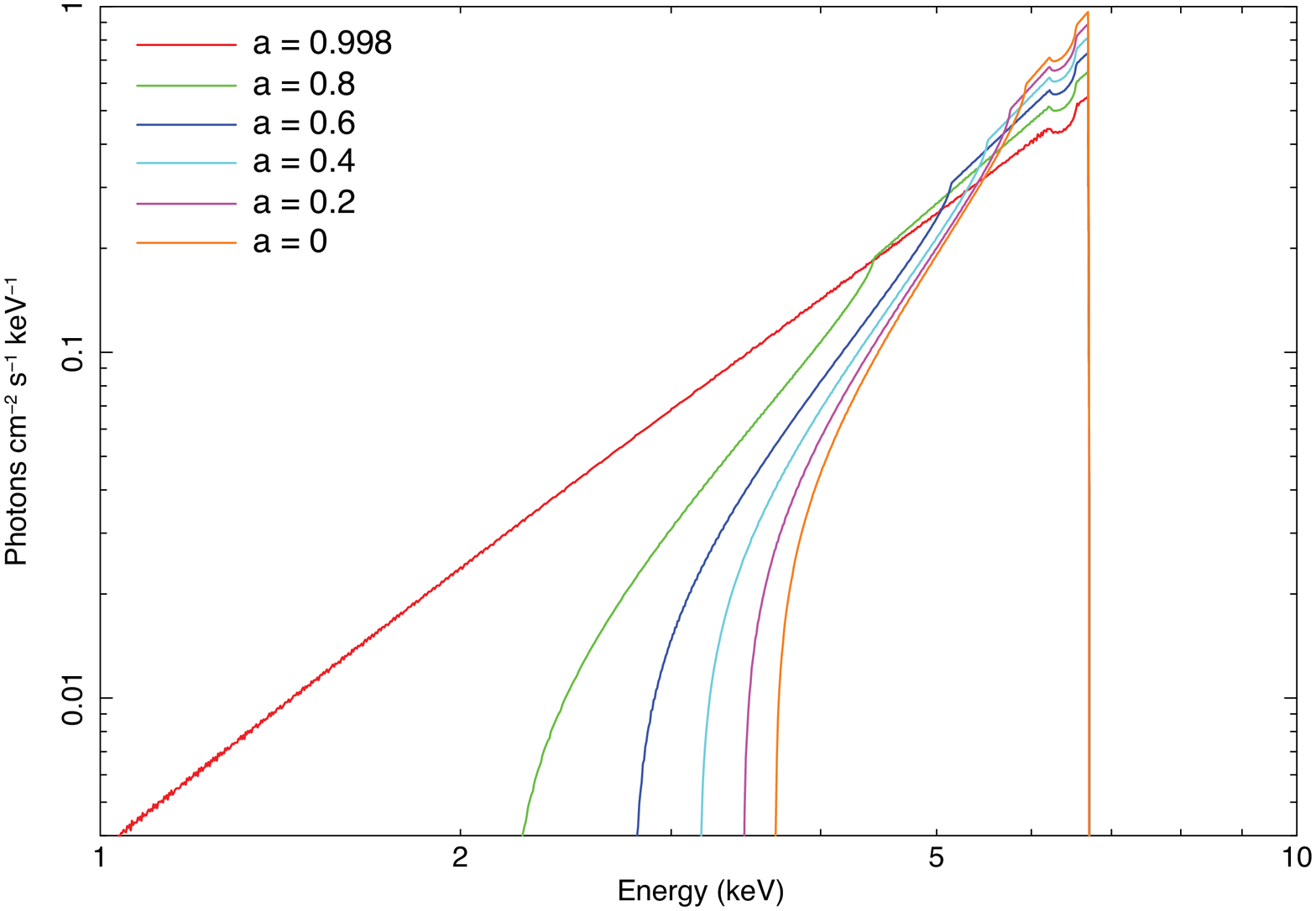}
\label{lines_spin.fig:full}
}
\subfigure[$10r_\mathrm{g} < r < 400r_\mathrm{g}$]{
\includegraphics[width=8cm]{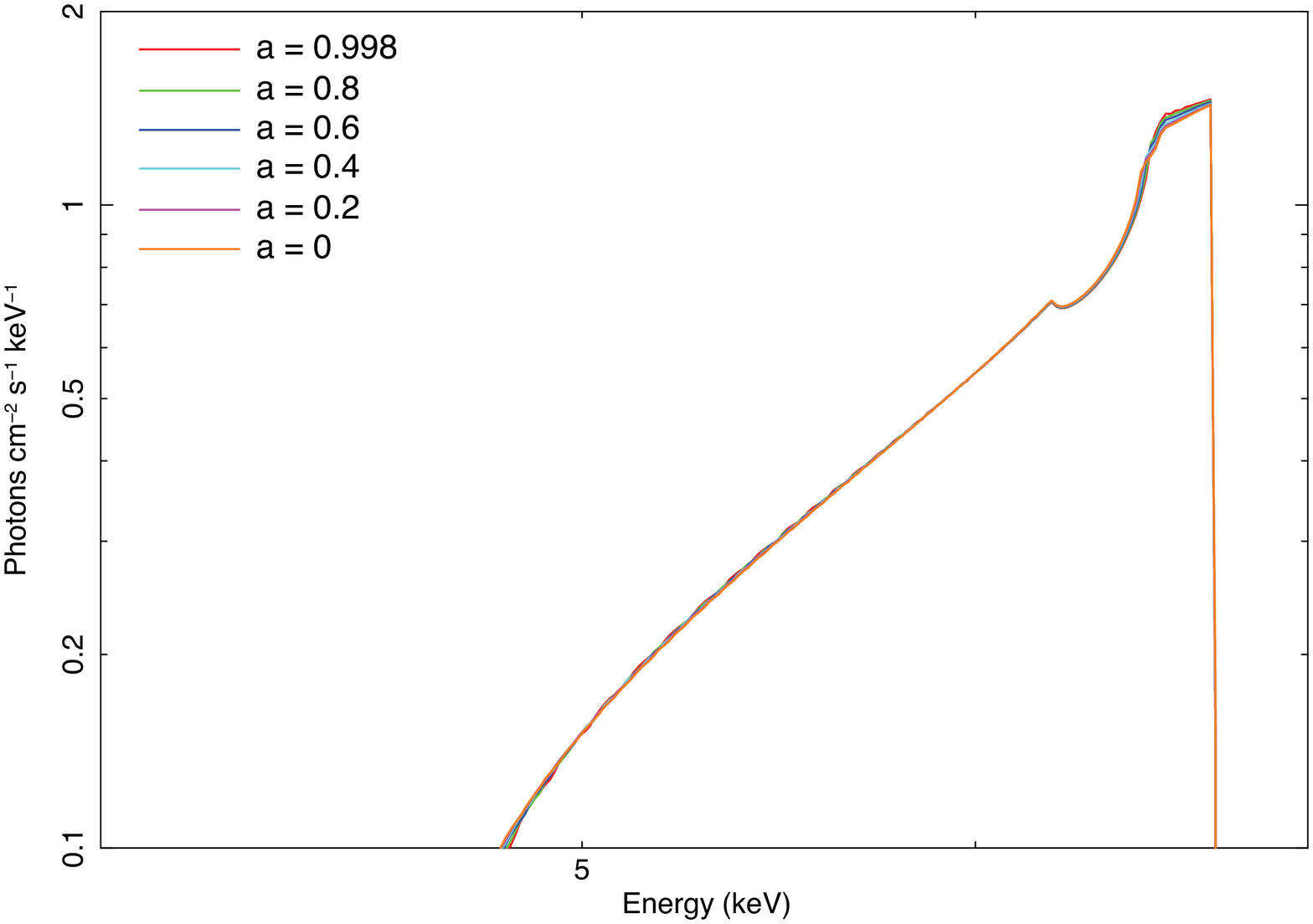}
\label{lines_spin.fig:10-400}
}
\subfigure[$10r_\mathrm{g} < r < 20r_\mathrm{g}$]{
\includegraphics[width=8cm]{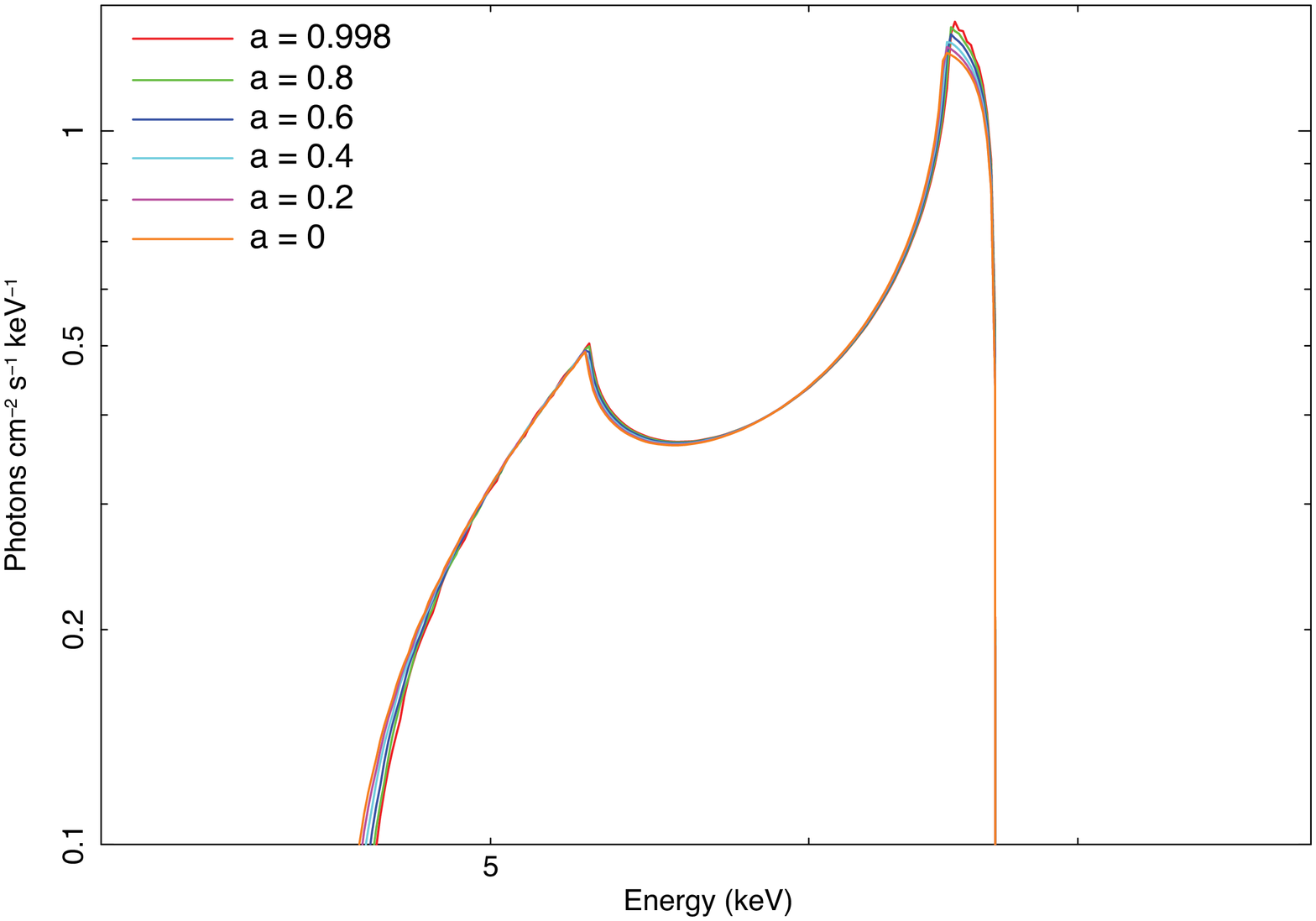}
\label{lines_spin.fig:10-20}
}
\caption[]{Theoretical line profiles for varying values of the black hole spin parameter, $a$, calculated using the \textsc{relline} model. \subref{lines_spin.fig:full} shows emission lines emitted from a region of the disc extending from the relevant innermost stable orbit (ISCO) for the value of $a$ out to 400\rg, with the extended redshifted wing of the line for higher values of $a$ resulting from the accretion disc (and thus emission) extending down to an ISCO closer to the black hole. \subref{lines_spin.fig:10-400} and \subref{lines_spin.fig:10-20}, showing lines emitted from the regions of the disc 10-400\rg\ and 10-20\rg\ respectively, demonstrate that the effect of the spin on the emission from a given radius is small. The effect of spin decreases in significance further from the black hole.}
\label{lines_spin.fig}
\end{figure}

While the profiles of emission lines originating from annuli in the accretion disc are calculated using the \textsc{laor} model and thus implicitly assume a maximally spinning Kerr metric ($a=0.998$), the profile of the line from a general annulus does not vary significantly with the spin parameter. This can be demonstrated by computing line profiles outwards of 6\rg\ (\textit{i.e.} the innermost stable orbit of a non-spinning, Schwarzschild black hole) for varying values of the spin parameter using a relativistic emission line model such as \textsc{relline} \citep{dauser+10}. The effect of varying the spin parameter is to move the innermost stable orbit, defining the innermost possible extent of the accretion disc and thus the extremal gravitational redshift, as illustrated in Figure \ref{lines_spin.fig}. As each annulus is considered independently, varying the black hole spin will determine whether contributions are detected from the inner annuli whose existence depends on the location of the ISCO. As such, the spin of the black hole will not affect the determined emissivity profile for the outer disc and the inference of emission from regions of the disc down to 1.235\rg\ implies a spin of the black hole close to maximal.

Theoretical emissivity profiles calculated here are for the simple case, considering only direct radiation from a point source. We have not in the present work considered the effect of spatially extended X-ray sources or the effect of radiation returning to the disc after reflection and hence being reflected for a second time \citep{cunningham-76} which may modify the final emissivity profile (for instance, increasing the reflection from the inner regions of the disc as more rays will be focussed to regions closer to the black hole, giving a steeper inner emissivity profile) and sharpen the spectrum \citep{ross+02}. This will be considered in future work.

It should be noted that if this twice-broken power law for the emissivity profile is averaged over the disc to a single-slope power law (averaging the index over radius, weighted by the annular area, $r\,dr$), the single power law index is found to be around 3.3. This would be found if a reflection component blurred using a single-slope emissivity profile is fit to the spectra and is only slightly steeper than the classical case, masking the significant steepening over the inner disc by relativistic effects.

Throughout the preceding analysis, the standard limb-darkening of emission from the accretion disc \citep{laor-91} is assumed. Work by \citet{svoboda+10} suggests that the disc emission due to reflection can in fact be limb-brightened or at least isotropic from the innermost regions of the disc close to the central black hole. This will lead to an enhanced red wing of the emission line over the typical limb-darkened profile as can be seen in the \textsc{relline} model of \citet{dauser+10}, and as such will lead to an apparently steeper emissivity profile. Thus, in reality the power law slope over the inner disc may be less than 7.9 in 1H\,0707-495, while the effect on the emissivity profile obtained will be less relevant for the outer disc regions where relativistic effects are weaker.

\section{Conclusions}

The X-ray reflection emissivity profile of the accretion disc in 1H\,0707-495 has been determined from the profile of the iron K$\alpha$ emission line in the X-ray spectrum without \textit{a priori} assumption of its form, by considering independently the relative contributions from successive radii in the disc.

The emissivity profile was found to agree with theoretical predictions; a steeply falling profile in the inner regions of the disc with a power law index of 7.8, then flattening to constant emissivity between 5\rg\ and 35\rg\ before tending to a constant power law of index 3.3 over the outer regions of the disc. The profile obtained suggests emission right down to the last stable orbit at 1.235\rg.

The emissivity profile obtained is consistent with fitting a relativistically blurred reflection spectrum from a disc with a continuous twice-broken power law emissivity profile, which provides a better fit than the previously assumed once-broken power law.

The observed profile can be understood in the context of gravitational light bending as rays travel around the black hole from the source to the disc before being reflected, as well as relativistic effects on the disc and agrees with forms predicted by ray tracing simulations, indicating that reflection from an accretion disc with a steep central emissivity profile is plausible.

The results can be compared to theoretical emissivity profiles computed for a range of locations and geometries of the X-ray source as well as disc geometries in order to place constraints on the properties of the coronal X-ray sources in AGN.

\bibliographystyle{mnras}
\bibliography{agn}

\begin{thebibliography}{}

\bibitem[\protect\citeauthoryear{{Cunningham}}{{Cunningham}}{1976}]{cunningham%
-76}
{Cunningham} C., 1976, \apj, 208, 534

\bibitem[\protect\citeauthoryear{{Dabrowski} et~al.}{{Dabrowski}
  et~al.}{1997}]{dabrowski+97}
{Dabrowski} Y., {Fabian} A.~C., {Iwasawa} K., {Lasenby} A.~N.,  {Reynolds}
  C.~S., 1997, \mnras, 288, L11

\bibitem[\protect\citeauthoryear{{Dabrowski} \& {Lasenby}}{{Dabrowski} \&
  {Lasenby}}{2001}]{dabrowski_lasenby}
{Dabrowski} Y.,  {Lasenby} A.~N., 2001, \mnras, 321, 605

\bibitem[\protect\citeauthoryear{{Dauser} et~al.}{{Dauser}
  et~al.}{2010}]{dauser+10}
{Dauser} T., {Wilms} J., {Reynolds} C.~S.,  {Brenneman} L.~W., 2010, \mnras,
  1460

\bibitem[\protect\citeauthoryear{{Fabian} et~al.}{{Fabian}
  et~al.}{1989}]{fabian+89}
{Fabian} A.~C., {Rees} M.~J., {Stella} L.,  {White} N.~E., 1989, \mnras, 238,
  729

\bibitem[\protect\citeauthoryear{{Fabian} et~al.}{{Fabian}
  et~al.}{2009}]{fabian+09}
{Fabian} A.~C. et~al., 2009, \nat, 459, 540

\bibitem[\protect\citeauthoryear{{George} \& {Fabian}}{{George} \&
  {Fabian}}{1991}]{george_fabian}
{George} I.~M.,  {Fabian} A.~C., 1991, \mnras, 249, 352

\bibitem[\protect\citeauthoryear{{Laor}}{{Laor}}{1991}]{laor-91}
{Laor} A., 1991, \apj, 376, 90

\bibitem[\protect\citeauthoryear{{Miniutti} et~al.}{{Miniutti}
  et~al.}{2003}]{miniutti+03}
{Miniutti} G., {Fabian} A.~C., {Goyder} R.,  {Lasenby} A.~N., 2003, \mnras,
  344, L22

\bibitem[\protect\citeauthoryear{{Ross} \& {Fabian}}{{Ross} \&
  {Fabian}}{2005}]{ross_fabian}
{Ross} R.~R.,  {Fabian} A.~C., 2005, \mnras, 358, 211

\bibitem[\protect\citeauthoryear{{Ross}, {Fabian}, \& {Ballantyne}}{{Ross}
  et~al.}{2002}]{ross+02}
{Ross} R.~R., {Fabian} A.~C.,  {Ballantyne} D.~R., 2002, \mnras, 336, 315

\bibitem[\protect\citeauthoryear{{Suebsuwong} et~al.}{{Suebsuwong}
  et~al.}{2006}]{suebsuwong+06}
{Suebsuwong} T., {Malzac} J., {Jourdain} E.,  {Marcowith} A., 2006, \aap, 453,
  773

\bibitem[\protect\citeauthoryear{{Svoboda} et~al.}{{Svoboda}
  et~al.}{2010}]{svoboda+10}
{Svoboda} J., {Dov{\v c}iak} M., {Goosmann} R.~W.,  {Karas} V., 2010, in
  American Institute of Physics Conference Series, Vol. 1248, {A.~Comastri,
  L.~Angelini, \& M.~Cappi} , ed, American Institute of Physics Conference
  Series, p. 515

\bibitem[\protect\citeauthoryear{{Zoghbi} et~al.}{{Zoghbi}
  et~al.}{2010}]{zoghbi+09}
{Zoghbi} A., {Fabian} A.~C., {Uttley} P., {Miniutti} G., {Gallo} L.~C.,
  {Reynolds} C.~S., {Miller} J.~M.,  {Ponti} G., 2010, \mnras, 401, 2419

\end{thebibliography}

\label{lastpage}

\end{document}